\documentclass[conference]{IEEEtran}

\usepackage{graphicx}
\usepackage{subfig}
\usepackage{algorithmic}
\usepackage{xcolor}
\usepackage{comment}
\usepackage{soul}
\usepackage{textcomp}
\usepackage{url}
\usepackage{amsmath} 

\AtBeginDocument{%
  }

\title{A Multi-Simulation Bridge for IoT Digital Twins}

\author{
    \IEEEauthorblockN{
        Marco Picone\IEEEauthorrefmark{1},
        Samuele Burattini\IEEEauthorrefmark{2},
        Marco Melloni\IEEEauthorrefmark{1},
        Prasad Talasila\IEEEauthorrefmark{3},
        Davide Ziglioli\IEEEauthorrefmark{1},\\
        Matteo Martinelli\IEEEauthorrefmark{1},
        Nicola Bicocchi\IEEEauthorrefmark{1},
        Peter Gorm Larsen\IEEEauthorrefmark{3}
    }

    \IEEEauthorblockA{
        \IEEEauthorrefmark{1}University of Modena and Reggio Emilia, Italy,
        \IEEEauthorrefmark{2}University of Bologna, Italy,\IEEEauthorrefmark{2}Aarhus University, Denmark\\
        \{name.surname\}@unimore.it, prasad.talasila@ece.au.dk, pgl@ece.au.dk,  samuele.burattini@unibo.it, a.ricci@unibo.it
    }
}

\begin{document}

\maketitle


\begin{abstract}
The increasing capabilities of Digital Twins (DTs) in the context of the Internet of Things (IoT) and Industrial IoT (IIoT) call for seamless integration with simulation platforms to support system design, validation, and real-time operation. This paper introduces the concept, design, and experimental evaluation of the \emph{DT Simulation Bridge}---a software framework that enables diverse interaction patterns between active DTs and simulation environments.
The framework supports both the DT development lifecycle and the incorporation of simulations during active operation. Through bidirectional data exchange, simulations can update DT models dynamically, while DTs provide real-time feedback to adapt simulation parameters. We describe the architectural design and core software components that ensure flexible interoperability and scalable deployment.
Experimental results show that the DT Simulation Bridge enhances design agility, facilitates virtual commissioning, and supports live behavioral analysis under realistic conditions, demonstrating its effectiveness across a range of industrial scenarios.
\end{abstract}

\section{Introduction}
\label{sec:introduction}

With the widespread adoption of Internet of Things (IoT) and Industrial IoT (IIoT) technologies, Digital Twins (DTs) have evolved into \emph{active} components---i.e., software entities that maintain continuous bidirectional communication with their physical assets~\cite{grieves2014digital, dt_manufacturing_review, dt_industry_statoftheart}.
This paradigm shift allows DTs to provide dynamic digital representations updated through sensor data streams, supporting real-time monitoring, anomaly detection, and adaptive control~\cite{Cimino_2019}. 

In this context, integrating simulations into the DT architecture becomes increasingly critical.
Simulations would enable DTs to perform ``what-if'' analyses, optimize performance parameters, and validate control strategies before applying them to the physical asset~\cite{Boschert2016}.
The combination of real-time data and virtual experimentation allows DTs to anticipate issues and adapt to changing conditions in complex environments.
More in detail, we identified four primary motivations for integrating simulations into DTs:

\begin{figure*}[ht]
    \centering
    \includegraphics[width=\textwidth]{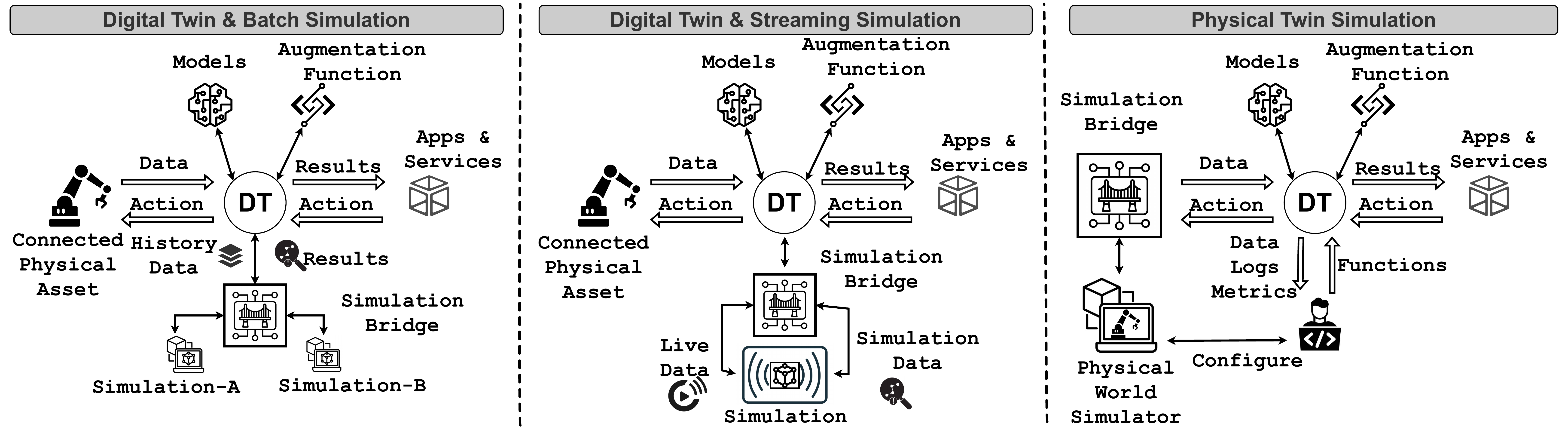}
    \caption{Overview of the Simulation Bridge enabling diverse interaction patterns between DTs and simulation environments.}
    \label{fig:dt_simulation_patterns}
    \vspace{-0.5cm}
\end{figure*}

\textit{(1) Enabling predictive and prescriptive capabilities.}  
While IoT-based DTs effectively capture the current state of physical systems through real-time data, open challenges remain in enhancing their ability to forecast future behavior and reason under uncertainty \cite{10015420}.
Simulation enhances DTs with predictive capabilities, supporting analysis under varying conditions such as equipment degradation, workload fluctuations, or environmental stressors~\cite{KLEIN2026103063}.
This shifts DTs from descriptive digital replicas to prescriptive and prognostic systems capable of anticipating outcomes and triggering proactive interventions.

\textit{(2) Coordinating heterogeneous simulations for complex systems.}  
Modeling complex systems often requires the orchestration of multiple, domain-specific simulations, each representing a different aspect of the physical system---e.g., mechanical dynamics, thermal effects, electrical behavior, or environmental interactions~\cite{10.1115/1.4049861,Fitzgerald2019}.
These simulations may operate at different time scales, resolutions, or levels of abstraction and must be integrated to ensure coherence and consistency. This is outside the scope of this paper.

\textit{(3) Supporting DT development in the absence of the Physical Twin (PT).}  
In many scenarios -- such as early-stage design, safety-critical applications, or high-cost manufacturing -- the physical system may not yet exist, be inaccessible, or pose operational risks.
In these cases, simulations -- e.g., of the IoT system~\cite{7568309} -- can act as surrogate PTs, generating mock behaviors and interaction patterns that allow the DT to be developed, tested, and validated in isolation~\cite{AITALLA20191331, 11087838}.

\textit{(4) Sharing simulation data across multiple DTs.}  
A single simulation model can serve multiple DTs, each potentially operating at different levels of abstraction and granularity.
For instance, a simulation of a production line may provide relevant data to DTs representing individual subsystems, such as robots or sensors, each requiring different views, resolutions, and update semantics. 

Despite these promising capabilities, the state of the art reflects only a partial integration of DTs and simulations~\cite{WOOLEY2023940}.
This paper presents the \emph{DT Simulation Bridge (DT-SB)}, a software framework designed to integrate DTs with multiple, heterogeneous simulation environments. The DT-SB enables bidirectional data exchange and supports diverse interaction patterns, allowing simulations to dynamically update the DT's internal state, while DTs can influence and tune simulation parameters in real time.
Unlike traditional co-simulation frameworks~\cite{Gomes&18}, the DT-SB connects a DT with multiple, independent simulations and acts solely as an intermediary. The coordination and aggregation of simulation results are delegated entirely to the DT, preserving architectural decoupling and maximizing flexibility.
We describe the modular architecture and core software components of the DT-SB, emphasizing its interoperability with diverse simulation models and protocols.
An experimental evaluation demonstrates the bridge’s effectiveness in accelerating iterative design cycles, facilitating virtual commissioning, and enabling live behavioral analysis under realistic operational conditions.


\section{Related Work}
\label{sec:relatedwork}

Simulation is a key technology in DTs, envisioned as a tool that can be adopted across all phases of the PT lifecycle~\cite{Boschert2016}.
Despite the strong overlap between simulation and DTs making it difficult to draw a line between the technologies~\cite{WOOLEY2023940}, it is becoming more and more clear that, especially in the IoT domain, DTs extend the capability of ``static'' simulations by incorporating real-time data~\cite{PHANDEN2021174}.

Most DT simulations, though, integrate simulators and IoT data processing with custom architectures and ad-hoc methods.
This lack of principled integration between the two worlds makes it hard to have a seamless experience especially when considering the possibility of using multiple simulation models within the same DT as we do in this paper.

The DT Simulation Bridge integrates and orchestrates multiple simulations with different models in a coherent framework.
At a surface level, it is then similar to distributed simulation~\cite{2000daglib} and co-simulation~\cite{Gomes&18}. 
There are, though, fundamental differences in the approach.
Distributed simulation usually concerns the execution of a single complex simulation on distributed nodes.
Differently, co-simulation explores the possibility of composing different simulation models of individual parts to simulate a complex system and coordinate the progress of time between the different independent simulation units.
The purpose of the DT Simulation Bridge is, instead, to connect one DT with multiple simulations.
%
The coordination and combination of such simulation results are expected to be managed by the DT and not by the bridge itself, which only acts as an intermediary.
Nevertheless, despite these differences in purposes, there are some significant aspects from these two related areas which influence the design of the bridge. 

Integrating simulations requires standardising the interaction across components.
For distributed simulation, the High-Level Architecture (HLA) defines a set of constraints to design interoperable federations of simulations which can be executed on a distributed runtime infrastructure~\cite{568652}.
Despite in our case simulations do not need to exchange data between each other, the need for a standardized interface is essential to allow the DT to seamlessly interact with multiple different simulations.
Similarly, in the co-simulation domain, the Functional Mockup Interface (FMI) is an industrial de facto standard \cite{FMIStandard2.0} to bridge different simulation models.
There, models of different nature may declare connection points to create input/output chains which combine the output of individual solvers.
We follow this principle in the design of the DT-SB to support simulation models which can be updated over time with the data that is captured by the DT itself. 
\section{Simulation Bridge Interaction Patterns}
\label{sec:dt_sim_bridge}

The integration of DTs with simulation environments enables a range of interaction patterns designed to support different phases of the system lifecycle and corresponding operational requirements. These interaction patterns are characterized by bidirectional data exchange and control flow, allowing DTs to incorporate predictive capabilities while simulations remain continuously informed by real-time contextual updates.
Building on the DT-SB architecture, we identify three primary interaction paradigms: \emph{Digital Twin \& Batch Simulation}, \emph{Digital Twin \& Streaming Simulation}, and \emph{Physical Twin Simulation}, as illustrated in Figure~\ref{fig:dt_simulation_patterns}. Each paradigm serves distinct objectives, encompassing design validation, what-if scenario analysis, live system adaptation, and virtual commissioning.
Together, they provide a flexible framework for coupling DTs with simulation processes across the development and operational spectrum.

\paragraph{Batch Simulation}
\label{ssec:dt-batch-simulation}

Leverages one-shot simulation runs to analyze scenarios or validate design decisions related to the DT and its physical counterpart. Acting as an intermediary, the DT-SB receives inputs from the DT, such as the current system state, historical data, or scenario configuration parameters and subsequently triggers one or more simulation runs accordingly. These batch simulations execute to completion producing results that the DT-SB forwards back to the DT.
The resulting simulation outcomes enable the DT to refine its models, develop predictive maintenance schedules, and recommend adjustments. This pattern is particularly well suited for use cases requiring comprehensive, offline evaluations, such as design validation and strategic scenario planning.

\paragraph{Streaming Simulation}
\label{ssec:dt-streaming-simulation}

Facilitates dynamic interactions between the DT and the simulation environment through continuous data exchange. In this mode, the DT streams high-frequency operational data to the DT-SB, which relays it to a connected simulation that dynamically adapts its model based on the incoming input. The simulation continuously produces output data that is streamed back to the DT through the DT-SB.
This interaction enables live behavioral analysis and adaptive control.
Furthermore, deviations between live DT data and simulation outputs can serve as early indicators of anomalies or faults, thereby enabling proactive intervention.
This pattern is particularly well suited for applications demanding immediate feedback and continuous system optimization.

\paragraph{Physical Twin Simulation}
\label{ssec:physical-twin-simulation}

Involves the use of a simulation environment that functions as a digital replica of the PT.
This interaction mode is particularly valuable for virtual commissioning, operator training, and control logic testing, where system behaviors can be evaluated in a safe, simulated setting prior to deployment in the real world.
In this pattern, the DT is unaware of the fact that it is communicating with a simulator and exchange data and control commands that are actuated in the simulated environment.
DT developers can configure simulation parameters through the DT-SB in order to test the behavior of the DT under varying conditions.
%
This pattern enables thorough testing and validation within a controlled virtual environment, significantly reducing the costs and risks associated with physical prototyping and on-site experimentation.

\begin{figure*}[h]
    \centering
    \subfloat[Simulation bridge architecture.]{
        \includegraphics[width=0.35\linewidth]{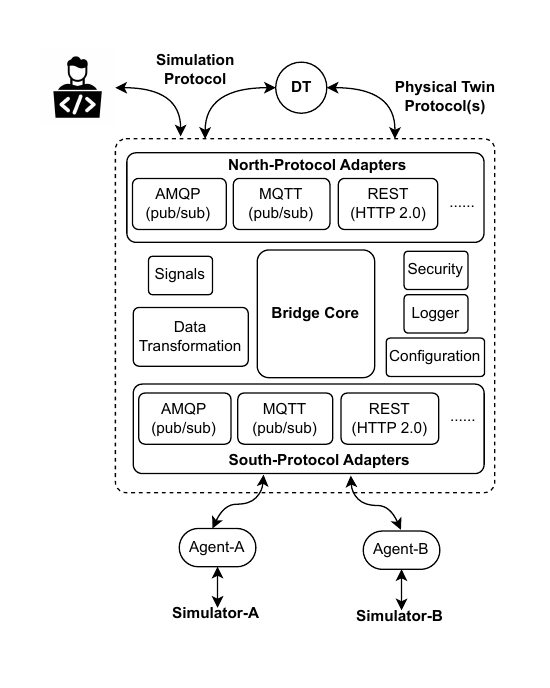}
        \label{fig:simulation_bridge_components}
    }
    \hfill
    \subfloat[Communication flow in the simulation bridge.]{
        \includegraphics[width=0.56\linewidth]{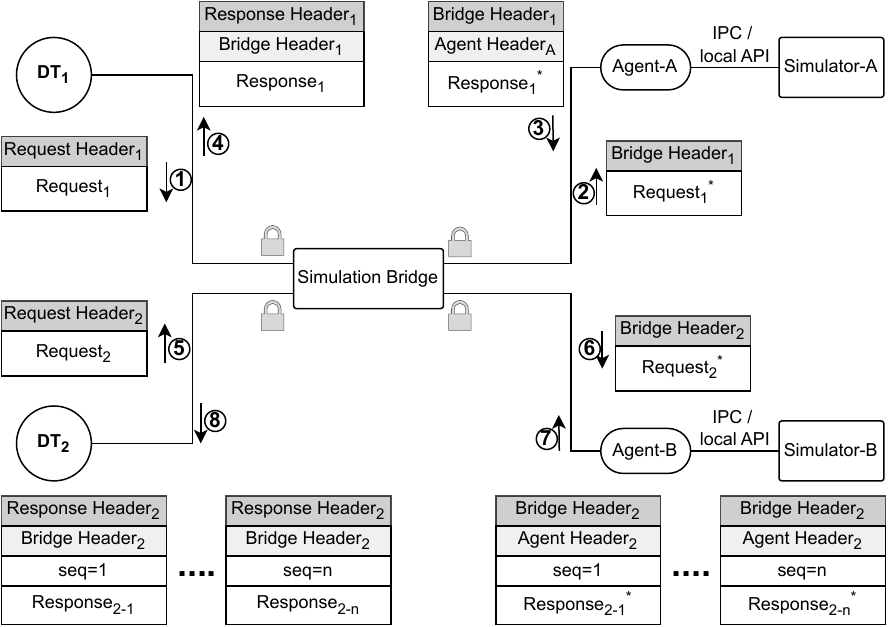}
        \label{fig:simulation_bridge_protocol}
    }
    \caption{Architecture of the Simulation Bridge with its main components and the associated interaction flows.}
    \label{fig:simulation_bridge_architecture}
    \vspace{-0.5cm}
\end{figure*}

\section{Simulation Bridge Architecture}
\label{sec:sim_bridge_arch}

The DT-SB coordinates heterogeneous and distributed simulations, providing a unified interface for DTs to access diverse simulation resources regardless of modeling and technological differences.
The bridge serves as a middleware which routes simulation execution requests, aggregates results, and returns them to clients. This enables seamless integration, reusability, and interoperability across simulation platforms.

Figure~\ref{fig:simulation_bridge_components} illustrates the modular architecture of the DT-SB. 
At the heart of this architecture is the DT-SB, each simulator is integrated through a dedicated simulation agent (SA) module.
SAs implement a standardized uniform API layer based on a simple application protocol. 
These agents are deployed within their respective simulation environments and are responsible for interpreting incoming messages, controlling the execution of simulation tasks, and aggregating the results for transmission back to the DT-SB.
%
Each module of the DT-SB is designed to fulfill a specific function within the operational workflow:

\noindent\textbf{Bridge Core:}  
Serves as the central control unit of the DT-SB. It coordinates interactions among internal modules, monitors communication flows, and orchestrates message routing between external clients (e.g., DTs, PTs, simulators) and the distributed simulation environment.

\noindent\textbf{Protocol Adapters (PAs):}  
facilitate protocol translation and connection management between the DT-SB and external entities. They operate on both the north-bound interface (connecting to DTs and PTs) and the south-bound interface (connecting to SAs). PAs are implemented using a plugin architecture, allowing the system to be extended in a modular fashion.

\noindent\textbf{Event Manager:}  
DT-SB adopts an event-driven architecture, wherein all relevant actions within a PA (e.g., message transmission, reception, connection management) are internally represented as events. When a request or response is received by a PA, it is transformed into an internal event and processed by the Bridge Core.

\noindent\textbf{Data Transformation Manager:} 
is responsible for converting data formats for both incoming and outgoing communication. Transformation parameters are dynamically configurable to optimize message structures for transmission efficiency. The module supports configurable message granularity, enabling the system to manage both atomic events and composite result packets based on application requirements.

\noindent\textbf{Configuration Manager:}  
handles the validation and loading of system parameters, including supported protocols, access credentials, and routing rules. A centralized configuration approach facilitates maintainability and control. The configuration also specifies a whitelist of authorized DTs and SAs.

\noindent\textbf{Logger:}  
records internal operations, exceptions, and diagnostic messages. It also provides performance metrics and operational insights essential for monitoring and debugging the DT-SB.

\noindent\textbf{Security Manager:}  
oversees the generation and loading of TLS certificates, ensuring secure communication channels. It is also responsible for detecting and mitigating malicious activity, such as message flooding, that could lead to denial-of-service conditions targeting simulators.

\subsection{Simulator Protocol}
\label{subsec:sim-protocol}

As DTs, DT-SB, and SAs operate in a distributed system, a dedicated \emph{Simulation Protocol} (SP) serves as the unifying layer for managing simulation requests and responses. The protocol defines standard headers for request routing and result interpretation. PAs integrated into DT-SB provide their own delivery semantics; however, best-effort delivery is universally supported, making it a practical foundation for interactions with SAs. Simulators can advertise the \emph{Simulation Type (ST)} they support (an abstraction that defines expected behaviors and model-method combinations). DTs may then select one of the available STs, while the DT-SB is responsible for routing and translating the corresponding simulation requests to the appropriate SAs that provide the selected ST.

The DT-SB performs two key operations to mediate this interaction:
(i) \emph{data transformation} of simulation requests and responses to ensure syntactic and semantic consistency, and
(ii) \emph{protocol header replacement} to enforce proper message routing between DTs and the selected SAs.
Figure~\ref{fig:simulation_bridge_protocol} depicts the message exchange flow for both batch and streaming simulation modes.
Secure variants of PAs are also illustrated. 
Numbered arrows indicate the sequence of message exchanges.

\paragraph{Batch Request}
In this scenario, $DT_1$ issues a simulation request targeting a Simulation Type supported by \emph{Simulator-A} (message $m_1$, denoted as \textcircled{1} in Figure~\ref{fig:simulation_bridge_protocol}). The request includes the model name, associated model parameters, simulator input values, and the specification of required outputs.
Upon receiving the request, the DT-SB replaces the original request header with a bridge-specific header. This header replacement ensures the necessary abstraction and transparency between DTs and SAs, decoupling clients from simulator-specific implementations. The DT-SB also performs a data transformation step to normalize the request format before forwarding it to \emph{Agent-A} (message $m_2$).
Agent-A then transmits the transformed request to \emph{Simulator-A}, which executes the simulation. The resulting outputs are sent back to the DT-SB via Agent-A (message $m_3$). Agent-A attaches a custom response header containing metadata relevant to the simulation run.
Upon receiving the response, the DT-SB applies a reverse data transformation tailored to the expectations of $DT_1$, replaces the agent-specific header with a standardized response header derived from the original request metadata, and delivers the final message to $DT_1$ (message $m_4$).
All headers in the message exchange carry metadata useful for distributed tracing, debugging, and performance analysis, thereby supporting observability and diagnostics across the system.

\paragraph{Streaming Request}
In this scenario, $DT_2$ initiates a simulation request targeting \emph{Simulator-B} (message $m_5$, denoted as \textcircled{5} in Figure~\ref{fig:simulation_bridge_protocol}). As with batch requests, the DT-SB intercepts the request, applies the necessary data transformation, and replaces the header with a bridge-specific version before forwarding the request to \emph{Agent-B} (message $m_6$). Agent-B then relays the request to \emph{Simulator-B}, which begins executing the streaming simulation.
Unlike batch simulations, streaming simulations generate a sequence of response messages over time. To ensure that responses are interpreted in the correct order, Agent-B attaches a sequence number to each outgoing message. This sequence number enables both DT-SB and $DT_2$ to preserve the temporal order of the response stream.
Agent-B forwards the ordered sequence of responses to the DT-SB (denoted as multiple instances of message $m_7$), which in turn applies reverse data transformation and forwards the processed responses to $DT_2$ (messages $m_8$).
In Figure~\ref{fig:simulation_bridge_protocol}, these streaming responses have $seq$ number field which is used by $DT_2$ to arrange the streaming responses in the correct order.

\paragraph{PT Simulation}

For the simulation of PTs, the DT-SB is used by two distinct entities. On one side, the developer interacts with the bridge through the simulation protocol and the north-protocol adapters to configure and launch simulations, define the mapping of simulated entities, and specify how these entities are exposed to the digital world via specific protocols, for example by mapping a simulation entity to MQTT. On the other side, the DT receives data from the simulated PTs through the configured protocols as if it were coming from the actual physical objects that will feed the DT in real deployments. Communication between the DT and the DT-SB is bidirectional: the DT not only receives telemetry data from the simulated PTs but can also send actions. These actions are translated by the bridge into commands for the simulation, modifying the behavior of the simulated PTs via interactions managed through the protocols and simulation agents enabled by the south-protocol adapters.

\subsection{Routing Messages}


A critical responsibility of the DT-SB is to map incoming simulation responses to their corresponding requests. This mapping is maintained using a dynamic routing table, with representative entries illustrated in Table~\ref{tab:index-table}. 

\begin{table}[h]
    \centering
    \caption{DT-SB's routing table.}
    \label{tab:index-table}
    \renewcommand{\arraystretch}{1.5}  
    \begin{tabular}{p{0.5in}llp{0.5in}lp{0.5in}}
    \hline
    \textbf{$PA_N$} & \textbf{$PA_S$} & \textbf{DT} & \textbf{Sim. Type} & \textbf{Request-id} & \textbf{Timeout}\\
    \hline
    MQTT & AMQP & $DT_1$ & Octave & $hash\#1$ & 60s\\
    AMQP & MQTT & $DT_1$ & Matlab & $hash\#2$ & 60s\\
    REST & AMQP & $DT_2$ & Simul8 & $hash\#3$ & 120s\\
    \hline
    \end{tabular}
\end{table}


The column $PA_N$ denotes the PA selected by the DT, while $PA_S$ represents the PA associated with the Simulator Agent. In scenarios where the DT-SB supports only a single protocol on either the north-bound ($PA_N$) or south-bound ($PA_S$) interface, the corresponding column can be omitted to simplify the routing logic.
The \emph{Simulator Type} column identifies the class of simulations handled by a given SA. Each simulation request is uniquely identified by a \texttt{RequestID}, which may be generated using hash-based algorithms to ensure uniqueness and unpredictability. A timeout field is also included to define the expiration threshold for requests; expired entries are purged to maintain routing table efficiency. When a simulation response arrives, the tuple $\langle PA_S, ST, \texttt{RequestID} \rangle$ is extracted and matched against existing routing table entries. The identified row provides the target DT and the associated $PA_N$, allowing the DT-SB to correctly route the response back to the originating DT. This mechanism ensures correct correlation of simulation data even in asynchronous or distributed execution contexts.

\section{Simulation Agents}
\label{sec:sim-agents}

This section presents the simulation agents integrated into the DT-SB architecture, focusing on the MATLAB and the AnyLogic Agents enabling the execution of computational and simulation models while maintaining modularity and seamless communication with the DT-SB. 

\subsection{Matlab Agent}

The MATLAB Agent ($SA_{\text{MATLAB}}$) serves as a decoupled simulation component integrating MATLAB’s computational capabilities within the DT-SB architecture. It communicates with the DT-SB through the AMQP protocol, enabling reliable and platform-independent message exchange. Four integration strategies were explored, each balancing latency, flexibility, and architectural coupling differently. 

In the \textit{file-based approach}, simulation inputs are provided as files, and results are written back to disk for the agent to read and forward to the DT-SB. This method ensures strong decoupling but introduces considerable input/output overhead, making it unsuitable for real-time or streaming simulations. The \textit{TCP/IP communication approach} enables real-time data exchange via socket-based interprocess communication, where results are transmitted as JSON messages. This solution supports continuous streaming and distributed deployment with moderate latency. A third strategy relies on shared memory \textit{interprocess communication}, where requests and responses are exchanged directly through memory segments. While this configuration achieves minimal latency and high throughput, it requires explicit synchronization and strict control over memory layout, reducing flexibility. Finally, the MATLAB Engine API allows direct invocation of MATLAB functions from Python, simplifying model execution but imposing tighter structural constraints and lowering modularity.

Experimental evaluation showed that shared memory communication provides the lowest latency but lacks adaptability. Therefore, the MATLAB Engine API was selected for \emph{batch} simulations, whereas the TCP/IP approach was adopted for \emph{streaming} mode, offering a balanced trade-off between performance, modularity, and ease of integration.

\subsection{AnyLogic Agent}

AnyLogic\footnote{AnyLogic: \url{https://www.anylogic.com/}}, is a flexible simulation platform that combines agent-based, discrete-event, and system dynamics modeling.
This versatility makes it ideal for creating detailed digital representations of industrial CPSs.
We hence develop the agent to support the PT simulation pattern. 

The AnyLogic Agent has two components:
1) an interface to the DT-SB which accepts simulation requests and configures the AnyLogic scenario accordingly. Differently from the other patterns, these requests come from the DT developer who wants to test the DT behaviour on a simulated PT;
2) a general purpose \emph{entity}\footnote{We use \emph{entity} here to avoid overloading the term agent. Since AnyLogic is an agent-based simulator, all entities within the simulations are modeled as agents which can exchange messages using built-in mechanisms.} that can be embedded within the simulation to receive messages from other entities and serialize them on a network socket to output a data stream of relevant events happening within the simulation. The simulation data stream is then received from the south protocol adapters, and routed to the physical interface of the DT with the north protocol adapters (Fig. \ref{fig:simulation_bridge_components}).

The AnyLogic data stream carries information about which simulation entity generated it, which can be used by the bridge to map the corresponding update on a different protocol for the DT.
This design is needed to emulate scenarios where data about one or more PTs is sent over different communication protocols (e.g. a sensor network), making it possible to test the DT simply by changing its configuration to point to the emulated communication channels and leaving all the connection handling and message processing logic unaltered.
The connection is bi-directional, to allow the DT to send control commands and apply them onto the simulated PT.

\begin{figure*}[h]
    \centering
    \subfloat[The UR10e robotic arm in MatLab.]{
        \includegraphics[width=0.465\linewidth]{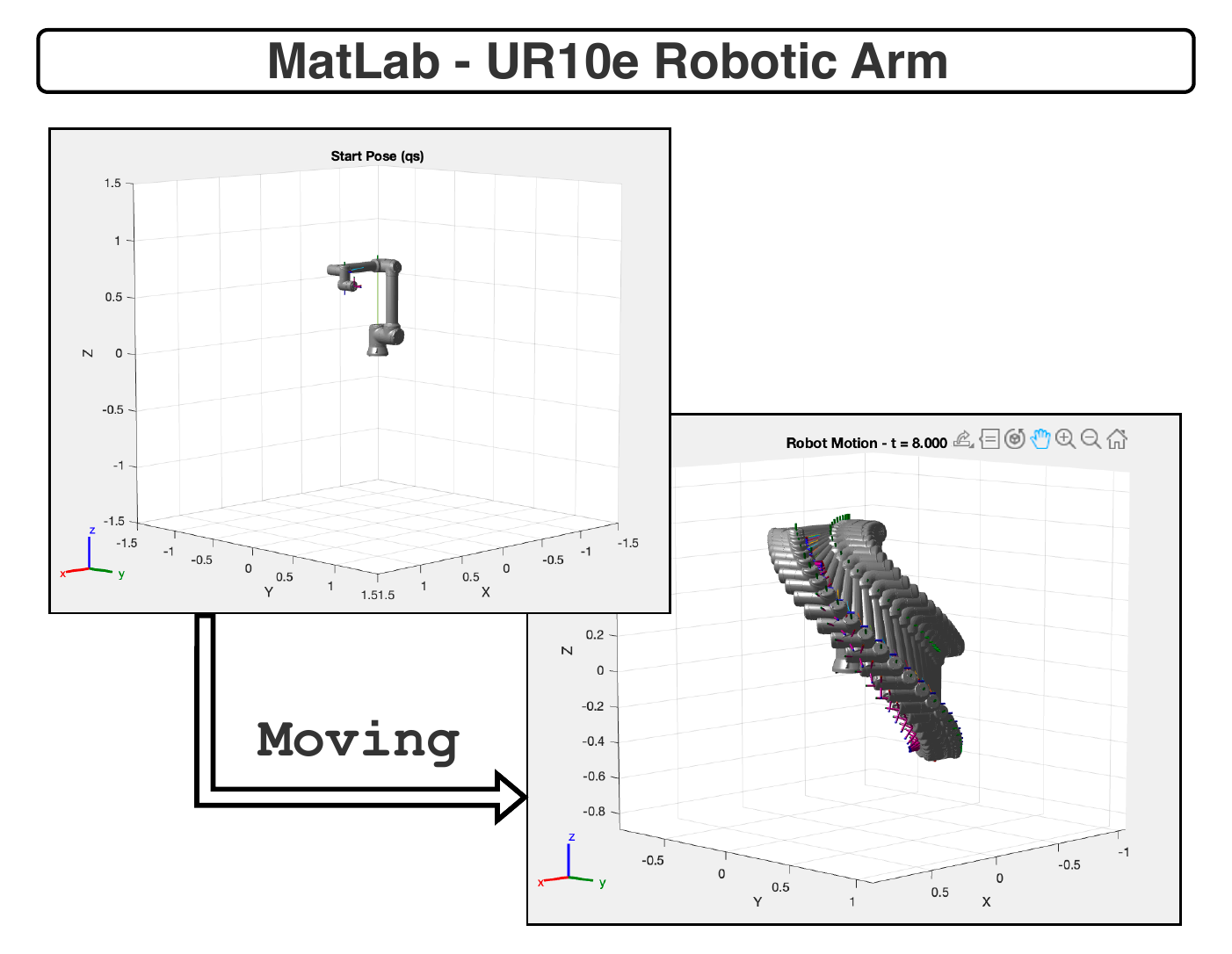}
        \label{fig:matlab_pt}
    }
    \hfill
    \subfloat[AnyLogic Microfactory simulation.]{
        \includegraphics[width=0.48\linewidth]{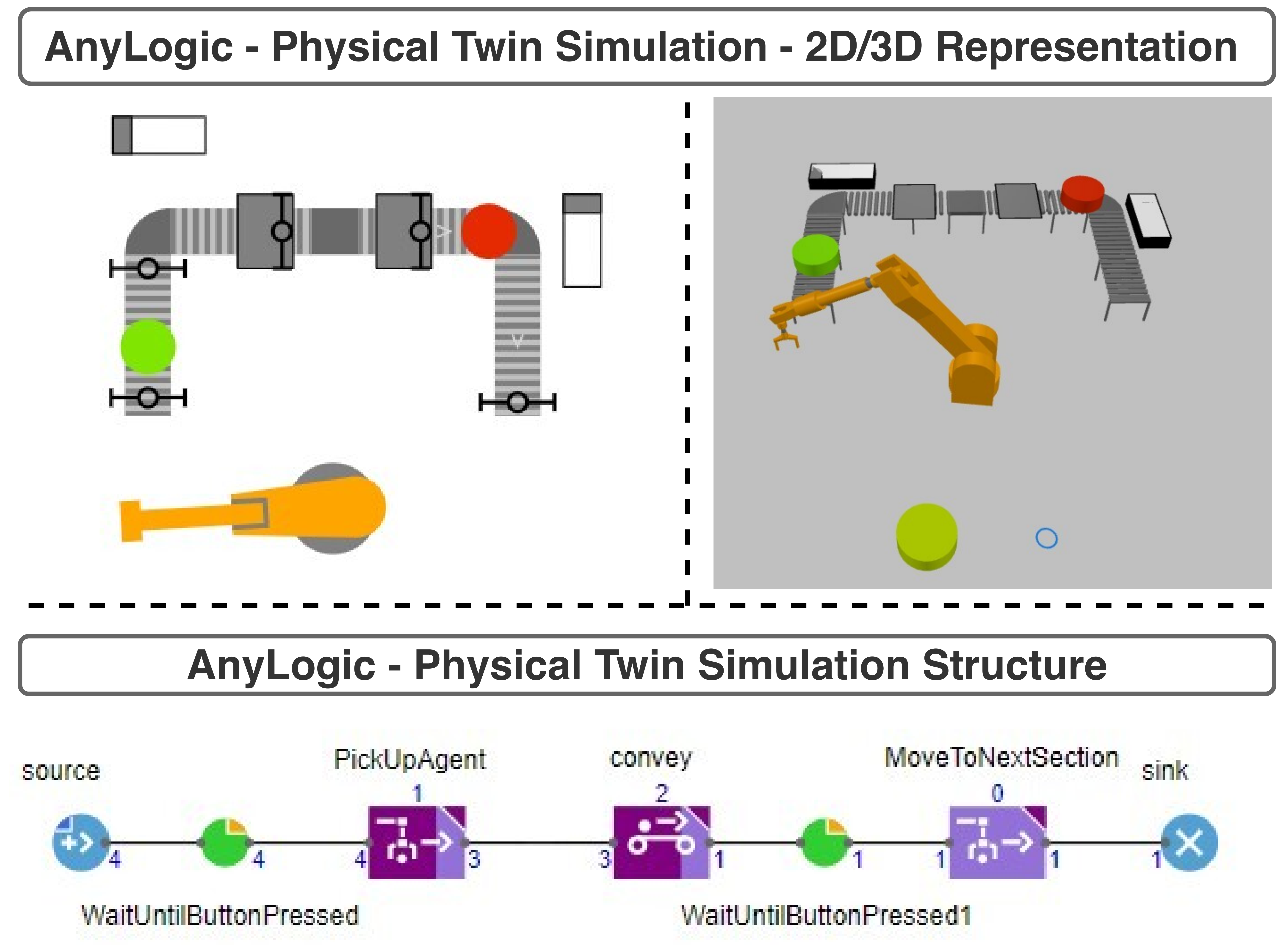}
        \label{fig:any_logic_simple}
    }
    \caption{Simulated Physical Twins integrated with the Simulation Bridge.}
    \label{fig:physical_twin_simulations}
    \vspace{-0.5cm}
\end{figure*}

\begin{table}[htbp]
\centering
\caption{Performance of DT-SB for different protocols.}
\label{tab:dt-sb_performance}
\renewcommand{\arraystretch}{1.5}
\begin{tabular}{p{0.3in}p{0.4in}p{0.3in}p{0.75in}p{0.75in}}
\hline
\textbf{Protocol} & \textbf{Overhead} & \textbf{STD} & \textbf{5\% Threshold} & \textbf{95\% Threshold} \\
\hline
MQTT & 2.47ms & 10.49ms & 1.69ms & 11.98ms \\
AMQP & 1.95ms & 0.99ms & 1.12ms & 3.69ms \\
REST & 4.63ms & 4.72ms & 2.92ms & 11.42ms \\
\hline
\end{tabular}
\vspace{-0.5cm}
\end{table}


\section{Experimental Evaluation}
\label{sec:experimental_evaluation}

To evaluate the performance of the proposed DT-SB and its integration with simulation agents, we conducted a series of experiments focusing on protocol overhead and agent-level performance on commercial computing hardware (MacBook Air Apple Silicon M2 processor and 8~GB of unified memory) with Python implementation of the proposed DT-SB \footnote{Simulation Bridge GitHub Repo: \url{https://github.com/INTO-CPS-Association/simulation-bridge}}. 


\subsection{Simulation Bridge Overhead}

\textbf{DT-SB Overhead}. The DT-SB was implemented to support multiple communication protocols on its north-bound interface (facing DTs), including MQTT, AMQP, and HTTP/2.0. Its south-bound interface (facing simulation agents) currently supports AMQP. We evaluated the processing time overhead introduced by the DT-SB core, explicitly excluding the contribution of individual protocol adapters.
Table \ref{tab:dt-sb_performance} shows the descriptive statistics of measured processing time overhead for MQTT, AMQP, and HTTP/2.0 protocol adapters.
These results confirm that DT-SB can reliably handle heterogeneous PAs with minimal latency in request forwarding and agent selection.

\textbf{Matlab Agent Overhead}.
To assess the performance of the MATLAB agent independently of simulation model complexity, we used lightweight test cases to isolate the agent's intrinsic processing overhead. Each simulation request was assigned a unique \textit{Request ID}, enabling precise tracking of request-response cycles. For streaming mode experiments, different time steps were used to vary the frequency of data exchange, simulating increased communication loads.
The overhead refers to the time spent by the agent to process a simulation request, excluding MATLAB startup time and simulation runtime. This metric captures the cost of request parsing, file operations, and inter-process communication. In batch mode, the agent exhibited a mean overhead of 2.7~ms. In streaming mode, the mean overhead was 3.7~ms. Table~\ref{tab:matlab-streaming} reports the measured overhead for different streaming frequencies, showing that agent performance remains stable across varying time steps. These findings confirm that $SA_{\text{MATLAB}}$ introduces minimal latency, making it suitable for both batch and real-time streaming applications.

\begin{table}[h!]
\centering
\caption{Matlab Agent's overhead for different frequencies.}
\label{tab:matlab-streaming}
\begin{tabular}{lcccccc}
\hline
\textbf{Streaming Frequency (ms)} & 10 & 50 & 70 & 100 & 150 & 200 \\ 
\hline
\textbf{Agent Overhead (ms)} & 2.2 & 4.1 & 5.6 & 4.9 & 3.2 & 3.0 \\ 
\hline
\end{tabular}
\end{table}

\textbf{Anylogic Agent Overhead}. To evaluate the performance of the AnyLogic agent a dedicated testing campaign was conducted to measure the system’s intrinsic overhead, independently of simulation model complexity aligned with the MATLAB agent analysis. The simulation used for the experiment is shown in Figure~\ref{fig:any_logic_simple}. The AnyLogic model employed is event-based, meaning that the simulation flow is internally controlled and does not rely on a fixed time step, unlike Table~\ref{tab:matlab-streaming}, which reports the MATLAB agent’s overhead at different frequencies. To ensure comparability with previous analyses, the simulation was executed at various run-time speeds (1x, 2x, 5x, 10x, 25x, 50x). The results show an extremely low and stable average overhead, ranging from 3.4~ms to 8.6~ms across different streaming configurations, as reported in Table~\ref{tab:anylogic-streaming}. These findings confirm that $SA_{\text{AnyLogic}}$ introduces negligible latency, making it suitable for real-time streaming applications.

\begin{table}[h!]
\centering
\caption{AnyLogic Agent's overhead at different execution speeds.}
\label{tab:anylogic-streaming}
\begin{tabular}{lcccccc}
\hline
\textbf{Execution Speed} & 1x & 2x & 5x & 10x & 25x & 50x \\ 
\hline
\textbf{Agent Overhead (ms)} & 8.6 & 4.9 & 8.3 & 3.4 & 5.3 & 3.6 \\ 
\hline
\end{tabular}
\end{table}

\subsection{Physical Twin Simulations}

To validate and analyze the proposed Simulation Bridge framework and the integration with DTs, two simulation environments were developed to emulate distinct aspects of the PTs behaviors: one in \textit{MATLAB/Simulink}, focusing on robotic motion control and kinematics, and one in \textit{AnyLogic}, modeling a microfactory scenario for system-level DT integration and communication assessment.

The first simulation involves a Universal Robots UR10e robotic arm executing a smooth point-to-point motion in joint space as depicted in Figure \ref{fig:matlab_pt}. The objective is to move the robot from an initial joint configuration to a target configuration within a specified time, following an optimally smooth trajectory and employing a simple closed-loop controller to track this trajectory. To ensure smooth motion, a fifth-order (quintic) polynomial trajectory was defined for each of the robot’s six joints. A quintic polynomial has six coefficients, allowing the imposition of six boundary conditions (initial and final position, velocity, and acceleration). This ensures continuity of the first and second derivatives—velocity and acceleration—throughout the entire motion.

The robotic simulation was implemented using the Robotics System Toolbox and a rigid-body model of the UR10e with gravity enabled. At each time step, the control law computed a new joint velocity command, and joint angles were updated via simple Euler integration. This iterative process continued until the planned trajectory duration elapsed. Completion was detected when the final joint configuration error dropped below a predefined negligible threshold, ensuring accurate convergence to the target configuration.

The second simulated environment was developed using the DT-SB integrated agent within \textit{AnyLogic} to support DT creation for a Microfactory scenario. The model represents an indexed production line comprising a conveyor, multiple workstations, and a robotic arm responsible for handling parts between loading and unloading areas, as illustrated in Figure~\ref{fig:any_logic_simple}. 


The two-dimensional model includes conveyors capable of reporting operational status, photocells detecting product passage, and transfer stations with custom switches emulating pusher mechanisms. Workstations are equipped with virtual actuators that communicate their operational state, while the central robot, modeled via the \textit{Robot} element, manages item positioning and state updates. Process logic is implemented using a block-diagram approach: agents are generated by a \textit{Source} block, controlled by a \textit{Hold} block linked to user interface commands, processed by a \textit{ProcessByRobot} block, transferred using a \textit{Convey} block, and finally removed by a \textit{Sink} block. 

Together, these simulation environments provided a controlled and complementary validation setup for the proposed approach. The MATLAB/Simulink model enabled low-level evaluation of motion control and trajectory tracking, while the AnyLogic microfactory scenario assessed system-level communication, synchronization, and interoperability. This combination demonstrated the framework’s flexibility in connecting diverse simulation domains within a unified DT integration architecture.

\section{Conclusion}
\label{sec:conclusion}

We propose the \emph{DT Simulation Bridge (DT-SB)}, a lightweight middleware architecture and protocol that unifies data exchange and interoperability between digital twins and heterogeneous simulation environments. The DT-SB enables seamless integration across \emph{batch}, \emph{streaming}, and \emph{PT simulation} interaction patterns, supporting diverse DT lifecycle applications. The DT-SB handles data translation, protocol adaptation, and synchronization between DTs and external simulators. Its effectiveness was demonstrated through two case studies based on MATLAB and AnyLogic simulations of cyber-physical scenarios. Future work will extend this approach toward co-simulation and hybrid setups, refined synchronization mechanisms, and adaptive orchestration in distributed environments together with the integration with DT open-source frameworks \cite{PICONE2021100661}.


\bibliographystyle{IEEEtran}
\bibliography{bibliography}

\end{document}